\documentclass[12pt,a4]{article}
\usepackage[utf8]{inputenc}
\usepackage{graphicx}
\usepackage{amsthm,amsmath,latexsym,amssymb,amsfonts,amscd}
\usepackage{graphics,lscape,fancyhdr,array,stmaryrd,euscript}
\usepackage{ifpdf}
\usepackage{verbatim,slashed}
\usepackage{relsize,setspace}
\numberwithin{equation}{section}
\usepackage{hyperref}

 \usepackage[numbers,sort&compress]{natbib}
 \usepackage[nottoc]{tocbibind}

\newcommand{\be}{\begin{equation}}
\newcommand{\ee}{\end{equation}}

\newcommand{\besubeqs}{\begin{subequations}}
\newcommand{\esubeqs}{\end{subequations}}

\begin{document}
\hfill
\begin{flushright}
    {LMU-ASC 37/17}
\end{flushright}
\vskip 0.02\textheight
\begin{center}

{\Large\bfseries AdS/CFT in Fractional Dimension \\ 

\vspace{0.4cm}
and Higher Spin Gravity at One Loop
\vspace{0.4cm}
} \\

\vskip 0.04\textheight

Evgeny \textsc{Skvortsov},${}^{1,2}$ Tung \textsc{Tran},${}^{1}$

\vskip 0.04\textheight

{\em ${}^{1}$ Arnold Sommerfeld Center for Theoretical Physics\\
Ludwig-Maximilians University Munich\\
Theresienstr. 37, D-80333 Munich, Germany}\\
\vspace*{5pt}
{\em ${}^{2}$ Lebedev Institute of Physics, \\
Leninsky ave. 53, 119991 Moscow, Russia}

\vskip 0.02\textheight

{\bf Abstract }

\end{center}
\begin{quotation}
Large-$N$, $\epsilon$-expansion or the conformal bootstrap allow one to make sense of some of conformal field theories in non-integer dimension, which suggests that AdS/CFT may also extend to fractional dimensions. It was shown recently that the sphere free energy and the $a$-anomaly coefficient of the free scalar field can be reproduced as a one-loop effect in the dual higher-spin theory in a number of integer dimensions. We extend this result to all integer and also to fractional dimensions. Upon changing the boundary conditions in the higher-spin theory the sphere free energy of the large-$N$ Wilson-Fisher CFT can also be reproduced from the higher-spin side. 
\end{quotation}

\newpage

\tableofcontents

\section{Introduction}
\label{sec:Intro}
There is some evidence that the $AdS/CFT$ correspondence  \cite{Maldacena:1997re,Witten:1998qj,Gubser:1998bc} may extend to fractional dimensions. Our goal is to support this idea by matching the  sphere free energy of the free and critical vector models with the one-loop corrections to the vacuum partition function in the higher-spin gravity. 

It was conjectured in \cite{Klebanov:2002ja} that the large-$N$ $O(N)$ vector model, which describes the critical points of $O(N)$ magnets in three dimensions \cite{Brezin:1972fb,Wilson:1973jj}, should be dual to a theory with gauge fields of arbitrary high spin in the spectrum, which are known as higher-spin theories.\footnote{See also \cite{Sundborg:2000wp,Sezgin:2002rt,Sezgin:2003pt,Leigh:2003gk} for other developments of this and related conjectures that involve higher-spin theories. } As it was mentioned already in \cite{Klebanov:2002ja}, see also \cite{Giombi:2014xxa}, the fact that the Wilson-Fisher critical point exists in $4-\epsilon$ expansion should allow one to make sense both of the dual higher-spin theory and of the duality itself in $AdS_{5-\epsilon}/CFT^{4-\epsilon}$. However, it is difficult at present to come up with an observable amenable to computation on the higher-spin side, especially in fractional dimension. In the paper we confirm the duality between free and critical large-$N$ $O(N)$ vector models at the one-loop level with the $AdS$ observable being the one-loop determinant of the Type-A higher-spin theory in Euclidean $AdS_{d+1}$ and with the dual CFT observable being the sphere free energy $F=-\log Z_{S^d}$. The result holds true in all integer and non-integer dimensions, which extends and generalizes  \cite{Giombi:2013fka,Giombi:2014iua}. Upon changing the boundary conditions we reproduce the difference between the sphere free energy under a double trace deformation $(\phi^2)^2$ that drives the free model at UV to the critical model in IR. 

The inspiration comes mostly from the CFT side, which is much better understood: there are different techniques available that allow one to make sense of at least some of the interacting conformal field theories in fractional dimensions. For example, the large-$N$ expansion, see e.g. \cite{Vasiliev:1981yc,Lang:1992zw,Petkou:1994ad,Moshe:2003xn}, can be used to compute the critical indices for any $d$, including non-integer ones. The large-$N$ approximation is also important for the quasi-classical expansion on the $AdS$-side since the coupling constant in higher-spin theories $G$ should be of order $1/N$. Another ubiquitous method is the $\epsilon$-expansion \cite{Wilson:1973jj}. Also, the conformal bootstrap can be set up in fractional space-time dimensions \cite{El-Showk:2013nia} and used to get predictions for the critical indices and to clearly show how the $2d$ Ising model smoothly turns into the $3d$ Ising model and ends up on the free theory in $4d$, the latter region can be accessed via the $4-\epsilon$ expansion too. The whole range $2<d<4$ is covered by the $1/N$-expansion whenever $N$ is large. There are recent studies \cite{Fei:2014yja,Mati:2016wjn} pointing out that the critical vector model can be extended to a wider range of dimension $4\leq d\leq6$.

An interesting class of observables that can be computed on both sides of the duality comes from the sphere free energy $F=-\log Z_{S^d}$. It is also related to an important problem of how to define a measure for the number of effective degrees of freedom in a general QFT. Such an observable should decrease along RG flow and be stationary at fixed points which are described by conformal field theories. The $d=2$ case is solved by the $c$-theorem \cite{Zamolodchikov:1986gt}, while the $4d$ case by the $a$-theorem \cite{Cardy:1988cwa,Komargodski:2011vj}. Both the central charge $c$ and the $a$ anomaly can be extracted from the sphere free energy: $F=a \log R$, where $R$ is the radius of the sphere and $c=-3a$ in $2d$. In $d=3$ there is no conformal anomaly but it was first conjectured \cite{Myers:2010xs,Jafferis:2011zi,Klebanov:2011gs} and then proved \cite{Casini:2011kv,Casini:2012ei} that $F$ works in $3d$ as well. More generally, $\tilde{F} =(-)^{({d-1})/{2}}\log Z_{S^d}$ is expected \cite{Klebanov:2011gs} to work in odd $d$, in particular in $d=1$ it gives the $g$-theorem \cite{Affleck:1991tk}. The last step \cite{Giombi:2014xxa} is to extend this definition to fractional dimensions by introducing generalized sphere free energy $\tilde{F} =\sin(\tfrac{\pi d}{2})\log Z_{S^d}$. This observable can be computed in non-integer dimension and the pole near even dimensions is resolved in such a way that the $a$-anomaly is captured, $\tilde F=(-1)^{d/2}\pi a/2$. For free CFT's or for the interactions induced by a double-trace deformation $\tilde{F}$ was computed in \cite{Gubser:2002vv,Diaz:2007an,Allais:2010qq,Klebanov:2011gs,Aros:2011iz}. For the free scalar field it is 
\begin{align}\label{eq:FA}
    \tilde F^\phi=\frac{1}{\Gamma (d+1)}\int_0^1 u \sin (\pi  u) \Gamma \left(\frac{d}{2}+u\right) \Gamma \left(\frac{d}{2}-u\right) \, du\,,
\end{align}
while for the change $\delta \tilde F$ induced by a double trace deformation due to an operator $O_\Delta$ of dimension $\Delta$ it is given by
\begin{align}\label{eq:deltaFA}
     \delta \tilde F_\Delta=\frac{1}{\Gamma (d+1)}\int_0^{\Delta-d/2} u \sin (\pi  u) \Gamma \left(\frac{d}{2}+u\right) \Gamma \left(\frac{d}{2}-u\right) \, du\,.
\end{align}
and we are interested in the case $\Delta=d-2$ that corresponds to $O=\phi^2$. 

On the dual AdS side the sphere free energy $F$ should correspond to the partition function in the Euclidean $AdS_{d+1}$, whose boundary is the sphere $S^d$. Optimistically, one should be able to match all the terms in the two expansions 
\begin{align}
    -\ln Z_{AdS}&=F_{AdS}=\frac{1}{G} F^0_{AdS} +F^1_{AdS} +G F^2_{AdS}+...\,,\\
    -\ln Z_{CFT}&=F_{CFT}=N F^0_{CFT}+F^{1}_{CFT}+\frac1{N}F^2_{CFT}+...\,.    
\end{align}
This idea was pursued in \cite{Giombi:2013fka} and elaborated further in \cite{Giombi:2014iua,Giombi:2014yra,Beccaria:2014xda,Beccaria:2014jxa,Beccaria:2014zma,Beccaria:2014qea,Basile:2014wua,Beccaria:2015vaa,Beccaria:2016tqy,Bae:2016rgm,Bae:2016hfy,Pang:2016ofv,Gunaydin:2016amv,Giombi:2016pvg,Bae:2016xmv,Brust:2016xif}. Generic duals of higher-spin theories are free CFT's as it is only in free CFT's one can have unbroken higher-spin symmetry \cite{Maldacena:2011jn,Alba:2013yda,Boulanger:2013zza,Stanev:2013qra,Alba:2015upa}. For free CFT's only the leading term $F^0_{CFT}$ is present, while it is not so for the large-$N$ interacting vector model. However, $F^0_{AdS}$ has not yet been computed since it should be equal to a regularized value of the classical action, which is not yet available. Still one can proceed to the one-loop term $F^1_{AdS}$ that is equal to the determinant $|-\nabla^2+m^2|$ of the kinetic terms of free higher-spin fields summed over an appropriate spectrum determined by the symmetry (or by the spectrum of the currents in the free CFT dual). This one-loop vacuum partition function, i.e. the one-loop determinant, can be computed via $\zeta$-function \cite{Dowker:1975tf,Hawking:1976ja} as 
\begin{align}
    F_{AdS}^1=-\frac12 \zeta'_{\Delta,s}(0)- \zeta_{\Delta,s}(0)\log l\Lambda
\end{align} 
for each individual field of spin-$s$ and dual operator dimension $\Delta$ and then summed up over a given spectrum (with the ghosts of the massless fields subtracted). We restrict ourselves to the simplest higher-spin theory, called Type-A, whose spectrum consists \cite{Vasiliev:2003ev} of totally symmetric massless fields with all integer spins $s=0,1,2,3,4,...$ (non-minimal Type-A) or all even spins $s=0,2,4,6,...$ (minimal Type-A). The minimal Type-A theory should be dual to free or critical $O(N)$ vector model while the non-minimal one should be dual to the $U(N)$ versions of the same CFT's. Whether the dual is free or interacting depends on the boundary conditions imposed on the scalar field, $s=0$, of the higher-spin multiplet: $\Delta=d-2$ for the free dual and $\Delta=2$ for the (large-$N$) interacting one. Therefore, altogether we have four different cases:\footnote{The second term in the brackets is to subtract the ghosts. }
\begin{align}\label{hszeta}
\begin{aligned}
    \zeta^{}_{\text{HS},\text{n.-m.}}(z)&=\zeta_{\Delta,0}+\sum_{s=1,2,...}\left[\zeta_{d+s-2,s}-\zeta_{d+s-1,s-1}\right]\,, \\
    \zeta^{}_{\text{HS},\text{min.}}(z)&=\zeta_{\Delta,0}+\sum_{s=2,4,...}\left[\zeta_{d+s-2,s}-\zeta_{d+s-1,s-1}\right]\,,
\end{aligned}
\end{align}
where $\Delta$ can be either $d-2$ or $2$. The one-loop free energy in the higher-spin gravity is
\begin{align}\label{fullfhs}
    F_{\text{HS}}^1=-\frac12 \zeta'_{\text{HS}}(0)- \zeta_{\text{HS}}(0)\log l\Lambda \,.
\end{align} 
It was shown in a number of integer dimensions \cite{Giombi:2013fka,Giombi:2014iua,Giombi:2014yra} that: (i) while each term in the sum may depend on the cutoff $\Lambda$, which makes the finite part ambiguous, the full one-loop vacuum energy does not depend on the cutoff $\Lambda$, i.e. $\zeta_{\text{HS}}(0)=0$ for the (non)-minimal Type-A models; (ii) the finite part vanishes for the non-minimal Type-A, $\zeta'_{\text{HS},\text{n.-m.}}(0)=0$, and equals the sphere free energy $F$ or the $a$-anomaly of the free scalar field, i.e. $a=-\tfrac12 \zeta'_{\text{HS},\text{min.}}(0)$ for $d$ even and $F=-\log Z_{S^d}=-\tfrac12 \zeta'_{\text{HS},\text{min.}}(0)$ for $d$ odd. This result can be consistent with the $AdS/CFT$ duality provided the one-loop effect compensates for the integer shift in the relation between the bulk coupling constant $G$ and the number of fields $N$ on the CFT side, $G^{-1}\sim N-1$ (provided that $F^0_{AdS}$ does match $F^0_{CFT}$).

The upshot of the one-loop computations in higher-spin theories is that the one-loop vacuum energy can reproduce the $a$-anomaly coefficient of the free scalar CFT in even dimensions and the sphere free energy in odd dimensions, which was shown for a number of dimensions. Upon changing the boundary conditions for the $s=0$ member it was also shown that the difference $-\tfrac12\delta \zeta'_{\text{HS}}(0)$, which is due to the scalar field, matches the sphere free energy of the large-$N$ interacting vector model in $d=3$ \cite{Giombi:2013fka} and $d=5$ \cite{Giombi:2014iua}.

It is worth stressing that the computations of $\zeta_{\text{HS}}$ are quite different for even and odd dimensions and the requirement for $d$ to be an integer has been a crucial one. Another common feature of the one-loop computations in higher-spin theories is that the finite result for $\zeta_{\text{HS}}$ is obtained after an appropriate regularization of the sum over spins. 

In the paper we revise the problem of one-loop computations in Type-A higher-spin theory aiming at the general proof for all integer dimensions and also for non-integer ones. We show that $\tfrac12 \sin(\tfrac{\pi d}{2})\zeta'_{\text{HS}}(0)$ for the minimal Type-A theory does reproduce the generalized sphere free energy \eqref{eq:FA} for all $d$. For $\Delta=2$ boundary conditions on the $s=0$ field the one-loop result matches the change in the sphere free energy \eqref{eq:deltaFA} due to the double-trace deformation induced by operator $(\phi^2)^2$ on the CFT side \cite{Klebanov:2011gs,Giombi:2013fka,Giombi:2014iua}. Since the result is a technical one let us briefly discus the main steps. First of all, thanks to Camporesi and Higuchi \cite{Camporesi:1994ga} there is a representation of the spectral density that enters $\zeta_{\Delta,s}(z)$ such that it can be extended to non-integer dimensions. Next, we apply the Laplace transform to the spectral density, see also \cite{Bae:2016rgm,Bae:2016hfy}, which disentangles the integral over the spectral parameter and summation over spins. Then, we convert the integral into a sum over the residues. In order to handle the sum we change the regularization prescription, see also \cite{Bae:2016rgm}, but it can be checked that this does not affect the result. Low and behold we arrive at the expression, which we refer to as {\it intermediate} form, whose regularized form gives \eqref{eq:FA}. The intermediate form can also be obtained directly on the CFT side from the determinant on the sphere. The interacting large-$N$ vector model requires taking into account the difference between the contributions of the $s=0$ fields for $\Delta=d-2$ and $\Delta=2$. Some other features and possible extensions are discussed in Conclusions.

\section{Higher-Spin Partition Function in Fractional Dimensions}
We first give in Section \ref{integerdimension} the basic definitions related to the zeta-function and review the computation of one-loop determinants in even and odd dimensions, stressing the difference. Next, in Section \ref{sec:TypeA}, we proceed to non-integer dimensions and apply the main technical tools that allow us to handle fractional dimensions: Laplace transform, contour integration and modified regularization. In Section \ref{sec:volume} we discuss the volume of the anti-de Sitter space that enters as an overall, but important, factor. The last steps on the AdS side --- summation over spins and extraction of $\zeta_{\text{HS}}(0)$ and $\zeta'_{\text{HS}}(0)$ are done in Sections \ref{sec:nonminA} and \ref{sec:minA}, where we arrive at certain {\it intermediate} form of the result that can be matched with the CFT side. The intermediate form is directly related to the free and critical vector models in Sections \ref{sec:matchingFree} and \ref{sec:matchingCritical}, which completes the proof.

\subsection{Integer Dimensions}
\label{integerdimension}
The general form of the spectral zeta-function in Euclidean anti-de Sitter space $AdS_{d+1}$ for a field of any symmetry type is \cite{Camporesi:1994ga}:
\begin{align}
    \zeta_{\Delta,s}&= \frac{\mathrm{vol}(\mathbb{H}^{d+1})}{\mathrm{vol}(S^d)} v_d g(s)\int_0^\infty d\lambda\, \frac{\mu(\lambda)}{\left[(\Delta -\tfrac{d}2)^2+\lambda ^2\right]^z}\label{zetageneral}\,,
\end{align}
where $\mu(\lambda)$ is the most important factor --- spectral density. It is normalized to its flat-space value:
\begin{align}
    \mu(\lambda)|_{\lambda\rightarrow\infty}&=w_d \lambda^{d}\,, &w_d&=\frac{\pi}{[2^{d-1}\Gamma(\tfrac{d+1}2)]^2}\,.
\end{align}
There are several overall factors that do not participate in the integral: $g(s)$ is the number of degrees of freedom, i.e. components of the irreducible transverse traceless tensor; $\mathrm{vol}(\mathbb{H}^{d+1})$ is a regularized volume of the hyperbolic space \cite{Diaz:2007an}; and the leftover factor $v_d$ can be combined with $w_d$:
\begin{align}
    v_d&= \frac{2^{d-1}}{\pi}\,, && u_d=v_dw_d=\frac{(\mathrm{vol}(S^d))^2}{(2 \pi )^{d+1}}\,.
\end{align}
We are interested in totally-symmetric massless spin-$s$ fields that make the spectrum of the Type-A higher-spin theory, in which case the number of degrees of freedom is 
\begin{align}\label{eq:spinfactorA}
    g(s)&=\frac{(d+2 s-2) \Gamma (d+s-2)}{\Gamma (d-1) \Gamma (s+1)}=\mathrm{dim}^{so(d)}\, \mathbb{Y}(s)\,.
\end{align}
It counts the dimension of an irreducible $so(d)$ tensor of rank-$s$. The spectral density depends crucially on whether $d$ is even or odd. For $d$ even, the spectral density is a polynomial:
\begin{align}\label{evendim}
    \mu(\lambda)&=w_d \left(\left(\frac{d-2}{2}+s\right)^2+\lambda ^2\right)\prod_{j=0}^{\frac{d-4}{2}} \left(j^2+\lambda ^2\right)\,,
\end{align}
while for $d$ odd it contains an additional $tanh$ factor:
\begin{align}\label{oddim}
     \mu(\lambda)&=w_d\lambda  \tanh (\pi  \lambda ) \left(\left(\frac{d-2}{2}+s\right)^2+\lambda ^2\right)\prod_{j=1/2}^{\frac{d-4}{2}} \left(j^2+\lambda ^2\right)\,.
\end{align}
The computation of $\zeta(z)$ in every even dimension $d$ presents no difficulty, in principle, since the spectral density is a polynomial and the integral can be done with the help of 
\begin{align}
    \int_0^\infty d\lambda\, \frac{\lambda ^k}{\left(b^2+\lambda ^2\right)^z}&= \frac{\Gamma \left(\frac{k+1}{2}\right) b^{k-2 z+1} \Gamma \left(-\frac{k}{2}+z-\frac{1}{2}\right)}{2 \Gamma (z)}\,.
\end{align}
Here, the spectral parameter $z$ serves also as a regulator. The only problem left is to evaluate the infinite sum over all spins \eqref{hszeta} and there are several equivalent ways known how to do that \cite{Giombi:2013fka,Giombi:2014iua,Giombi:2014yra}. Summation over spins can be done dimension by dimension with the complexity rapidly increasing with $d$.

On contrary, the computation in every odd dimension is a challenge. The coefficient of the $\log$ divergent piece, i.e. $\zeta_{\Delta,s}(0)$ can still be computed for any spin and weight $\Delta$ after splitting the integrand into the part that is simple and converges for $z$ large enough and another part that is more complicated but converges for $z=0$, which is done with the help of
\begin{align}
    \tanh x&=1+\frac{-2}{1+e^x}\,.
\end{align}
Computation of $\zeta'_{\Delta,s}(0)$ leads to some integrals that cannot be evaluated analytically, but whose contribution cancels out after summing over the spectrum of the Type-A theory. Again the computation can be done dimension by dimension. We refer to \cite{Giombi:2013fka,Giombi:2014iua} for more detail, see also \cite{Gunaydin:2016amv,Giombi:2016pvg}.

Another issue that requires a separate treatment is the factor of the regularized volume of the Hyperbolic space $\mathbb{H}^{d+1}$, which via dimensional regularization can be found to be \cite{Diaz:2007an}:
\begin{align}\label{eq:volumeinteger}
    \mathrm{vol}\, S^d&=\frac{2 \pi ^{(d+1)/2}}{\Gamma \left(\frac{d+1}{2}\right)}\,, &&
    \mathrm{vol}\, \mathbb{H}^{d+1}=
        \begin{cases}
        \frac{2 (-\pi )^{d/2} }{\Gamma \left(\frac{d}{2}+1\right)}\log R\,, & d=2k\,,\\
        \pi ^{d/2} \Gamma \left(-\frac{d}{2}\right)\,, & d=2k+1\,.
        \end{cases}
\end{align}
The appearance of $\log R$ signals conformal anomaly. The sphere free energy also has the $\log R$ term, whose coefficient is the $a$-anomaly.

\subsection{Fractional Dimensions}
\label{sec:TypeA}
Coming to fractional dimensions we prefer to isolate all the factors, including the volume of the hyperbolic space, and denote the leftover as $\tilde{\mu}(\lambda)$
\begin{align}\label{mathcalN}
    \zeta(z)&= \mathcal{N} g(s)\int_{0}^{\infty}d\lambda\,\frac{\tilde{\mu}(\lambda)}{\left[\lambda^2+\left(\Delta-\tfrac{d}{2}\right)^2\right]^z}\,, && \mathcal{N}=\frac{v_d w_d \mathrm{vol}(\mathbb{H}^{d+1})}{\mathrm{vol}(S^d)}\,.
\end{align}
There is a representation of the spectral density that works in all dimensions \cite{Camporesi:1994ga}:
\begin{align}\label{campor}
\tilde{\mu}(\lambda)&=\left(\left(\frac{d-2}{2}+s\right)^2+\lambda ^2\right)\left|\frac{\Gamma \left(\frac{d-2}{2}+i \lambda \right)}{\Gamma (i \lambda )}\right|^2\,.
\end{align}
This is our starting point. We will show that the higher-spin zeta-function can be computed without having to make an assumption that $d$ is an integer. The difference between even and odd dimensions can be observed in \eqref{campor}: a simple polynomial is obtained for $d$ even, \eqref{evendim}, and an additional non-polynomial factor is present for $d$ odd, \eqref{oddim}. In fact, the spectral density is not a polynomial in all dimensions, including fractional ones, except for the case of $d$ even. Therefore, it is the case of $d$ even that is special, all other dimensions being on equal footing. The computation we perform below is valid for all $d$ except even (which is of measure zero on the real line). The result for $d$ even is then obtained as a continuation from non-integer $d$.

Let us begin with the expression for the zeta-function that is obtained by collecting all the factors and expanding the gamma functions:
\begin{equation}\label{eq:zetaA}
    \zeta_{\nu,s}(z) = \mathcal{N}\frac{g(s)}{\pi}\int_0^{\infty} d\lambda \frac{\lambda \sinh(\pi \lambda)\left(\lambda^2+\left(\frac{d}{2}+s-1\right)^2\right) \Gamma\left(\frac{d}{2}+i\lambda -1\right)\Gamma\left(\frac{d}{2}-i\lambda -1\right)}{(\lambda^2+\nu^2)^z}\,,
\end{equation}
where $\nu=\Delta-\tfrac{d}{2}$. The integrand is an even function of $\lambda$ and therefore we can extend the range of integration to $(-\infty,\infty)$ at the price of $\tfrac12$. It is convenient to perform the Laplace transform, see also \cite{Bae:2016rgm,Bae:2016xmv},\footnote{One can represent the spectral zeta-function as a differential operator acting on some seed function that has enough parameters to produce $g(s)\mu(\lambda)$. Character is an example of such a function \cite{Bae:2016rgm,Bae:2016xmv}, which is also indispensable for taking tensor products. The characters are however difficult to define in non-integer dimension.} 
\begin{equation}\label{laplace}
    \frac{1}{(\lambda^2+\nu^2)^z}=\frac{\sqrt{\pi}}{\Gamma(z)}\int_0^{\infty} d\beta\, e^{-\beta \nu} \left(\frac{\beta}{2\lambda}\right)^{z-\frac{1}{2}}  J_{z-\frac{1}{2}}(\lambda \beta)\,.
\end{equation}
The main advantage is that the exponential $e^{-\beta \nu}$ times $g(s)$ can be summed over all spins in the spectrum directly. In other words, the sum over spins and the $\lambda$ integral are now decoupled. This is one of the crucial steps that allows us to calculate the full zeta function $\zeta_{\text{HS}}$, \eqref{hszeta}, in arbitrary dimension. Notice that in applying the Laplace transform we moved the branch point from $\pm i\nu$ in (\ref{eq:zetaA}) to $0$. Next, we split the Bessel function into 
\begin{equation}\label{Besselsplit}
    J_{\alpha}(x)=\frac{_1H_{\alpha}(x)\, +\, _2H_{\alpha}(x)}{2}\,,
\end{equation}
where $_1H_{\alpha}(x)$ and $_2H_{\alpha}(x)$ are Hankel functions of the first kind and second kind. 

\begin{figure}
    \centering
    \includegraphics[scale=0.4]{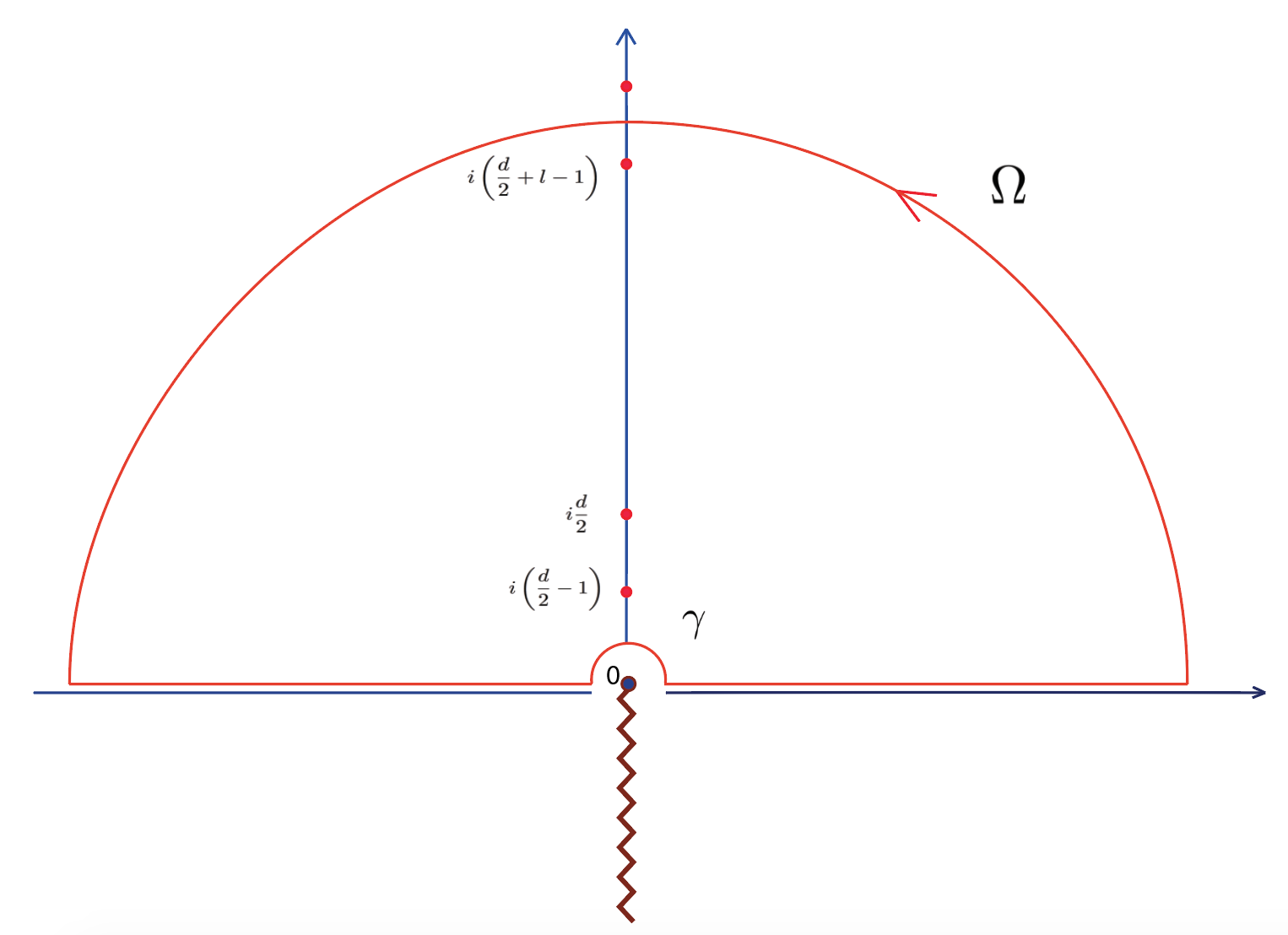}
    \caption{The contour for the part contains $_1H_{\alpha}$ lies in upper half plane where the poles are those of $\Gamma\left(\frac{d}{2}+i\lambda-1\right)$. As the $\lambda$ integral approaches $(-\infty,\infty)$, the range of $l$ also extends to infinity.}
    \label{fig:contour}
\end{figure}

Similarly to Green functions we close the contour for the part of $_1H_{\alpha}$ upward and the contour for the part of $_2H_{\alpha}$ downward. Let us show how to compute the contour integral of the part with $_1H_{\alpha}$ in (\ref{Besselsplit}) first. In order to evaluate the contribution coming from $_1H_{\alpha}$, we choose the contour as on Fig.~\ref{fig:contour}.
One needs to make sure that the upper arc of the contour does not cross any pole that comes from the $\Gamma\left(\frac{d}{2}+i\lambda-1\right)$. The residue theorem implies that
\begin{equation}\label{eq:res1}
    \oint_C f(\lambda) = 2\pi i\sum_{l=0}^{\infty} \text{Res}_{\lambda \rightarrow i\left(\frac{d}{2}+l-1\right)} \left(\lambda - i\left(\frac{d}{2}+l-1\right)\right) f(\lambda)\,,
\end{equation}
where we prefer to omit $g(s)/\pi$ for a moment:
\begin{equation}\label{eq:integrand1}
    f(\lambda)=\mathcal{N}\frac{\lambda \sinh(\pi \lambda)\left(\lambda^2+\left(\frac{d}{2}+s-1\right)^2\right) \Gamma\left(\frac{d}{2}+i\lambda -1\right)\Gamma\left(\frac{d}{2}-i\lambda -1\right) \beta^{z-\frac{1}{2}}\, _1H_{z-\frac{1}{2}}(\beta \lambda)}{2 (2\lambda)^{z-\frac{1}{2}}}
\end{equation}
We recall that the residues of $\Gamma$-function are 
\begin{equation}
 \text{Res}(\Gamma,-l) = \frac{(-1)^l}{\Gamma(l+1)}\,.
\end{equation}
We, therefore, could change the integral over $\lambda$ to an infinite sum over $l$. Before proceeding further, let us make sure that the upper arc and the contour around the branch point do not contribute to the whole contour integral. We make the change of variable $\lambda=R e^{i \theta}$:
\begin{equation}\label{eq:contour1}
    \int_{\Omega} d\lambda f(\lambda) = \lim_{R\rightarrow \infty} \int_0^{\pi} d\theta f(Re^{i\theta}) \qquad \text{and}\qquad  \int_{\gamma} d\lambda f(\lambda) = \lim_{R \rightarrow 0} \int_{\pi}^0 d\theta f(Re^{i\theta})\,.
\end{equation}
Introducing $z$ as a regulator \cite{Camporesi:1994ga,Giombi:2013fka,Giombi:2014iua,Giombi:2014yra} is useful in various ways. Let us consider the $\gamma$ contour first, if we set $z$ large enough then there is no contribution from the small contour
\begin{equation}\label{eq:bessel1}
   \lim_{\lambda \rightarrow 0} \lambda \sinh(\pi \lambda) \Gamma\left(\frac{d}{2}+i \lambda -1\right) \Gamma\left(\frac{d}{2}-i \lambda -1\right) \frac{\beta^{z-\frac{1}{2}}\, _1H_{z-\frac{1}{2}}(\beta \lambda)}{2(2\lambda)^{z-\frac{1}{2}}} = 0+ \mathcal{O}(\lambda^2) 
\end{equation}
Therefore, (\ref{eq:integrand1}) vanishes and the integral over the contour near the branch point in (\ref{eq:contour1}) also vanishes. Next, consider the large arc $\Omega$, assuming that the contour goes in between the poles of the gamma function. The integrand (\ref{eq:integrand1}) will also vanish as we make $z$ large enough in the limit where the radius $R$ goes to infinity.\footnote{For a more detailed discussion see \cite{Camporesi:1994ga}} Therefore, there is no contribution coming from $\gamma$ and $\Omega$ arcs and  (\ref{eq:res1}) is equal to
\begin{equation}
\begin{split}
    \frac{1}{2}\int_{-\infty}^{\infty}d\lambda f(\lambda) = & -\frac{\pi \mathcal{N}}{2} \sum_{l=0}^{\infty} \left(\frac{\beta^{z-\frac{1}{2}}\, _1H_{z-\frac{1}{2}}\left(i\left(\frac{d}{2}+l-1\right) \beta\right)}{\left(2 i\left(\frac{d}{2}+l-1\right)\right)^{z-\frac{1}{2}}}\right) 
    \left(\frac{d}{2}+l-1\right)\\ \times&\left(\left(\frac{d}{2}+s-1\right)^2-\left(\frac{d}{2}+l-1\right)^2\right)
    \sin\left(\pi \left(\frac{d}{2}+l-1\right)\right) \frac{\Gamma(d+l-2)}{\Gamma(l+1)} (-1)^l\,.
    \end{split}
\end{equation}
We notice that $\sin\left(\pi \left(\frac{d}{2}+l-1\right)\right) = - (-1)^l \sin\left(\frac{\pi d}{2}\right)$ and hence
\begin{equation}\label{expibeta}
\begin{split}
    \zeta(z) = & \frac{\mathcal{N}g(s)\sqrt{\pi}}{2\Gamma(z)} \int_0^{\infty} d\beta  \sum_{l=0}^{\infty} e^{-\beta \nu} \left(\frac{\beta^{z-\frac{1}{2}}\, _1H_{z-\frac{1}{2}}\left(i\left(\frac{d}{2}+l-1\right) \beta\right)}{\left(2 i\left(\frac{d}{2}+l-1\right)\right)^{z-\frac{1}{2}}}\right) 
    \\
    \times&\left(\frac{d}{2}+l-1\right)\left(\left(\frac{d}{2}+s-1\right)^2-\left(\frac{d}{2}+l-1\right)^2\right)\sin\left(\frac{\pi d}{2}\right) \frac{\Gamma(d+l-2)}{\Gamma(l+1)} \,.
    \end{split}
\end{equation}
It is difficult to say anything about the sum in general, but eventually we are interested only in few terms around $z=0$. In \cite{Bae:2016rgm}, it was argued that one can change the regularization prescription so that the $z\rightarrow0$ behaviour is not modified. Indeed, it is clear that to the leading order in $z$-expansion one can use
\begin{equation}\label{H1}
   \lim_{z\rightarrow 0}\frac{\beta^{z-\frac{1}{2}}\, _1H_{z-\frac{1}{2}}\left( \beta\lambda \right)}{2\left(2 \lambda\right)^{z-\frac{1}{2}}}= \frac{e^{i\beta\lambda}}{\sqrt{\pi}\beta}+\mathcal{O}(z)\,.
\end{equation}
This way we obtain the following contribution coming from the $_1H_{\alpha}$ function with the contour in the upper half-plane, Fig.~\ref{fig:contour}:
\begin{equation}\label{uppercontour}
\begin{split}
    \hat{\zeta}_{_1H_{\alpha}}(z) = & \frac{\mathcal{N}g(s)\sqrt{\pi}}{\Gamma(z)} \int_0^{\infty} d\beta  \sum_{l=0}^{\infty} e^{-\beta \nu} \frac{e^{-\beta\left(\frac{d}{2}+l-1\right)}}{\sqrt{\pi}\beta}
    \\
    \times&\left(\frac{d}{2}+l-1\right)\left(\left(\frac{d}{2}+s-1\right)^2-\left(\frac{d}{2}+l-1\right)^2\right)\sin\left(\frac{\pi d}{2}\right) \frac{\Gamma(d+l-2)}{\Gamma(l+1)} \,.
    \end{split}
\end{equation}
The presence of $1/\Gamma(z) \sim z$ factor in \eqref{laplace} implies that in order to get the right $\zeta(0)$ we can take only the constant term of \eqref{H1} into account. However, there should be a discrepancy between $\zeta'(0)$ computed rigorously and the one after we drop the term $\mathcal{O}(z) $ in \eqref{H1}. The difference, which we call the {\it deficit}, originates from the term of order $\mathcal{O}(z)$ in \eqref{H1}. As was noted in \cite{Bae:2016rgm} the deficit vanishes for representations that have even characters (even as a function of $\beta$, where $q=e^{-\beta}$ counts the energy via insertion of $q^E$). The deficit is discussed in Appendix \ref{app:modified}, where it is shown that it does not contribute to the full $\zeta'_{\text{HS}}(0)$.

Next, we repeat the same steps for the contribution coming from $_2H_{\alpha}$ in (\ref{Besselsplit}) where we close the contour downwards. In this case, one has to use $-2\pi i$ when applying residue theorem\footnote{The contour is the reflection image of Fig.~\ref{fig:contour} around the real axis.} for the poles of $\Gamma\left(\frac{d}{2}-i\lambda-1\right)$. 
We obtain the same structure as in (\ref{uppercontour}) since
\begin{equation}
    \lim_{\lambda \rightarrow -i(\frac{d}{2}+l-1)}   \lim_{z\rightarrow0}\, \frac{\beta^{z-1/2}\, _2H_{z-\frac{1}{2}}\left( \beta\lambda \right)}{2\left(2 \lambda)\right)^{z-\frac{1}{2}}}= \frac{e^{-\beta\left(\frac{d}{2}+l-1\right)}}{\sqrt{\pi}\beta}+\mathcal{O}(z)\,.
\end{equation}
Therefore, in order to compute the full one-loop free energy of the Type-A theory, \eqref{fullfhs}, we can write the zeta-function in a modified form as
\begin{equation}\label{template}
\begin{split}
    \tilde{\zeta}(z) &= \frac{\mathcal{N}g(s)}{\Gamma(2z)} \int_0^{\infty} d\beta  \sum_{l=0}^{\infty} e^{-\beta \nu} \beta^{2z-1}e^{-\beta\left(\frac{d}{2}+l-1\right)} \left(\frac{d}{2}+l-1\right) \\
    & \times \left(\left(\frac{d}{2}+s-1\right)^2-\left(\frac{d}{2}+l-1\right)^2\right) \sin\left(\frac{\pi d}{2} \right) \frac{\Gamma(d+l-2)}{\Gamma(l+1)}\,.
    \end{split}
\end{equation}

\subsection{Volume of Hyperbolic Space}
\label{sec:volume}
In integer dimensions, we can use the volume form of the sphere $S^{d}$ and Hyperbolic space as in (\ref{eq:volumeinteger}). This result arises \cite{Diaz:2007an} from the expansion of the formal volume $\pi ^{D/2} \Gamma \left(-\frac{D}{2}\right)$ in $D=d-\epsilon$:
\begin{align}\label{genvol}
  \mathrm{vol}\, \mathbb{H}^{d+1}=\frac{L_{d+1}}{\epsilon}+ V_{d+1}+ O(\epsilon)  \,,
\end{align}
where $\epsilon$-pole signals the $\log R$ divergence in $d=2k$ and $V_{d+1}$ is the finite part that makes the leading contribution for $d=2k+1$. As it was already noted in \cite{Diaz:2007an},  regularization of the volume IR divergences is not independent of regularization of the UV divergences that arise in one-loop determinants. Below, we propose an extension for the overall normalization factor which comes from the regularized volume to non-integer dimension. Note that one can write the general volume for Lobachevsky space as
\begin{equation}\label{volH}
    \text{vol} \, \mathbb{H}^{d+1}=-\frac{\pi^{\frac{d+2}{2}}}{\Gamma\left(\frac{d+2}{2}\right)\sin\left(\frac{\pi d}{2}\right)}\,,
\end{equation}
which gives the right pole as in \eqref{genvol} and reduces to $V_{d+1}$ for $d$ odd. The $\sin\left(\frac{\pi d }{2}\right)$ factor inside the modified zeta function (\ref{template}) will cancel with the one in (\ref{volH}) and gives us no poles for even dimensions. Together with the factor $\mathcal{N}$ in (\ref{mathcalN}), one arrives at the overall normalization factor in general dimensions\footnote{Recall that $\mathrm{vol}\, S^d=\frac{2 \pi ^{(d+1)/2}}{\Gamma \left(\frac{d+1}{2}\right)}$.}
\begin{equation}\label{overall}
    \widetilde{\mathcal{N}}=\mathcal{N}\sin\left(\frac{\pi d }{2}\right)=-\frac{1}{\Gamma(d+1)}\,.
\end{equation}
This overall normalization factor is strikingly simple since we do not need to treat the cases of odd and even dimensions separately. Moreover, (\ref{overall}) can also be used in fractional dimension.

\subsection{Non-minimal Type-A}
\label{sec:nonminA}
Using the regularized volume, we can now write the full modified zeta-function for the Type-A as
\begin{equation}\label{eq:modzetaA}
\begin{split}
    \tilde{\zeta}(z)_{\nu,s}=&-\frac{g(s)}{\Gamma(2z)\Gamma(d+1)}\int_0^{\infty}d \beta \sum_{l=0}^{\infty} e^{-\beta \nu}\beta^{2z-1} e^{-\beta\left(\frac{d}{2}+l-1\right)}\left(\frac{d}{2}+l-1\right)\\
    \times&\left(\left(\frac{d}{2}+s-1\right)^2-\left(\frac{d}{2}+l-1\right)^2\right) \frac{\Gamma(d+l-2)}{\Gamma(l+1)} 
    \end{split}
\end{equation}
We first show that the modified zeta-function leads to $\zeta_{\text{HS}}(0)=\zeta'_{\text{HS}}(0)$ for the non-minimal Type-A theory. The total $\zeta$-function for the non-minimal Type-A is 
\begin{equation}\label{eq:nonminimalA}
    \tilde\zeta_{\text{n.-m.}}(z) = \tilde\zeta_{\frac{d}{2}-2,0}(z)+\sum_{s=1}^{\infty} \left(\tilde\zeta_{\frac{d}{2}+s-2,s}(z)-\tilde\zeta_{\frac{d}{2}+s-1,s-1}(z) \right)\,,
\end{equation}
where the labels of the zeta functions correspond to $\zeta_{\nu,s}$ as in \eqref{eq:zetaA}.
Using (\ref{eq:modzetaA}) and (\ref{eq:spinfactorA}) we can perform the sum over all spins in \eqref{eq:nonminimalA} and get 
\begin{equation}\label{summeds}
\begin{split}
    \tilde\zeta_{\text{n.-m.}}(z)&=\sum_{l=0}^{\infty}\int_0^{\infty}\frac{d\beta\beta^{2z-1}}{\Gamma(d+1)\Gamma(2z)}\frac{e^{\frac{-\beta d}{2}}(-2+d+2l)\cosh\left(\frac{\beta}{2}\right)^2e^{-\frac{\beta}{2}(-2+d+2l)}\Gamma(-2+d+l)}{(1-e^{-\beta})^d \Gamma(l+1)}\\
    &\times (d^2+2(-2+l)l+d(-1+2l)-2l(-2+d+l)\cosh(\beta)) \\
    &=0\,.
    \end{split}
\end{equation}
It is the sum over $l$ that makes the expression  in (\ref{summeds}) vanish. Next, we need to compute $\tilde{\zeta}'_{\text{n.-m.}}(0)$ using the modified zeta-function. Remember that
\begin{equation}\label{eq:reduction1}
   \lim_{z\rightarrow 0}\frac{\beta^{2z-1}}{\Gamma(2z)}\sim \frac{2z}{\beta} + \mathcal{O}(z^2)\,.
\end{equation} 
In other words, the part of \eqref{summeds} without $1/\Gamma(2z)$ is $\tilde{\zeta}'(0)$. For the non-minimal Type-A we see that $\tilde{\zeta}'(0)$ vanishes. As a result we have proved that
\begin{equation}
    \tilde{\zeta}_{\text{n.-m.}}(0)=\tilde{\zeta}'_{\text{n.-m.}}(0)=0\,.
\end{equation}
This extends the results of \cite{Giombi:2013fka,Giombi:2014iua} to all odd dimensions as well as to fractional ones. 

\subsection{Minimal Type-A}\label{sec:minA}
The case of the minimal Type-A model is more interesting as we will not always find $0=0$-type of equality as in the non-minimal case. The $\zeta$-function for the minimal Type-A is
\begin{equation}\label{eq:sumeven}
    \zeta_{\text{min.}}(z)=\zeta_{\frac{d}{2}-2,0}(z)+\sum_{s=2,4,...}^{\infty} \left(\zeta_{\frac{d}{2}+s-2,s}(z)-\zeta_{\frac{d}{2}+s-1,s-1}(z) \right)\,.
\end{equation}
The final result after the summation is done has a very simple form:
\begin{equation}\label{eq:important}
    \tilde{\zeta}_{\text{min.}}(z)=-\frac{1}{2\Gamma(2z)}\int_0^{\infty}d\beta \frac{\beta^{2z-1}e^{-\beta(2-d)}(1+e^{2\beta})^2}{(e^{2\beta}-1)^d}\,.
\end{equation}
To obtain (\ref{eq:important}), it is suggestive to sum over $s$ in (\ref{eq:modzetaA}) first. To do this we need to absorb all monomials in $s$ into gamma functions. For example,
\begin{equation}
    s \Gamma(d+s-2)=\Gamma(d+s-1)-(d-2)\Gamma(d+s-2)
\end{equation}
After some algebra what we obtain are several terms of the form
\begin{equation}\label{eq:block}
    \xi(\nu,p(s))=e^{-\beta \nu} \frac{\Gamma(p(s))}{\Gamma(s+1)}\,.
\end{equation}
Here $p(s)$ is of the form $s+\text{const}$ with different $\text{const}$. The sums are of the usual "statistical" form. Following (\ref{eq:sumeven}) one should sum (\ref{eq:block}) according to
\begin{equation}
    \xi\left(\frac{d}{2}-2,p(0)\right)+\sum_{s=2,4,...}^{\infty} \xi\left(\frac{d}{2}+s-2,p(s)\right)-\xi\left(\frac{d}{2}+s-1,p(s-1)\right)\,,
\end{equation}
where  $\xi\left(\frac{d}{2}+s-1,p(s-1)\right)$ correspond to the ghosts. We, then, arrive at the sum over $l$:
\begin{equation*}
\begin{split}
\tilde{\zeta}_{\text{min.}}(z)&=\small\sum_{l=0}^{\infty} \int_0^{\infty}d\beta \frac{e^{-\beta\left(\frac{d}{2}+l-1\right)}(-d+2l-2)\Gamma(d+l-2)}{\Gamma(d+1)\Gamma(l+1)}\frac{e^{\beta(1-\frac{d}{2})}(-1+\coth\beta)\sinh\frac{\beta}{2}}{(1-e^{-2\beta})^d}\\
& \times \frac{\beta^{2z-1}}{\Gamma(2z)}\Bigg[-2(1+e^{-\beta})^d \left(\cosh\frac{\beta}{2}\right)^3\left((-1+d)d+2(-2+d)l+2l^2-2l(-2+d+l)\cosh\beta\right) \\
&+\cosh\beta((-1+d)d+2(-2+d)l+2l^2+2l(-2+l+1)\cosh\beta)\sinh \frac{\beta}{2}\left(1-e^{-\beta}\right)^d \Bigg]\\
    &=-\frac{1}{2\Gamma(2z)}\int_0^{\infty}d\beta \frac{\beta^{2z-1}e^{-\beta(2-d)}(1+e^{2\beta})^2}{(e^{2\beta}-1)^d}=\eqref{eq:important}\,.
    \end{split}
\end{equation*}
Formula (\ref{eq:important}) is strikingly simple. Vanishing of $\tilde{\zeta}_{\text{min.}}(0)$ is due to the fact that $\lim_{z\rightarrow0}1/\Gamma(2z)=0+\mathcal{O}(z)$. For $\tilde{\zeta}'_{\text{min.}}(0)$, using (\ref{eq:reduction1}), we arrive at 
\begin{equation}\label{eq:compactA}
    \tilde{\zeta}'_{\text{min.}}(0)=-\int_0^{\infty}d\beta \frac{e^{-\beta(2-d)}(1+e^{2\beta})^2}{\beta(e^{2\beta}-1)^d}\,. \qquad \quad 
\end{equation}
The formula above is the {\it intermediate}\footnote{We refer to it as intermediate as the integral is divergent and requires regularization.} form. After a suitable regularization it will give the correct answer for the sphere free energy as we recall in the next Section. It is worth mentioning that some of the intermediate, usually divergent, expressions on the $AdS$ side can be directly matched with their CFT cousins, see e.g. \cite{Giombi:2014yra} for the Casimir Energy example. These facts accentuate the importance of careful adjustment of the regularization prescriptions on both sides of the duality.

\subsection{Matching Free Vector Model}\label{sec:matchingFree}
Having arrived at the intermediate form \eqref{eq:compactA}, we would like to show that exactly the same intermediate form emerges on the CFT side. It contains all the important information and can be directly used to derive the sphere free energy. 

Let us review the main steps in \cite{Klebanov:2011gs,Giombi:2015haa,Giombi:2014xxa} as to get the (generalized) sphere free energy $\tilde F$. The starting point is the expression for $ {F}$ for a free scalar field, which results from the sum over the eigen values of the Laplace operator on the sphere $S^d$ \cite{Diaz:2007an,Gubser:2002vv}:
\begin{equation}\label{eq:loggamma}
    F^{\phi}_{\text{min.}}= \frac{1}{2}\sum_{l=0}^{\infty} d_l \log\frac{\Gamma(\frac{d}{2}+l-1)}{\Gamma(\frac{d}{2}+l+1)}= \frac{1}{2}\sum_{l=0}^{\infty} d_l \int_0^{\infty} \frac{d\beta}{\beta}\left(-2e^{-\beta}+e^{-\beta\left(l+\frac{d}{2}\right)}+e^{-\beta\left(\frac{d}{2}+l-1\right)}\right)\,,
\end{equation}
where
\begin{equation}
    d_l=\frac{(d+2l-1)\Gamma(d+l-1)}{\Gamma(d)\Gamma(l+1)}
\end{equation}
is the degeneracy of eigen values. There is a clearly divergent part proportional to the total number of 'degrees of freedom', $\sum d_l$. This sum can be shown to vanish in a number of ways. For example, inserting cut-off $e^{-\epsilon l}$ we get
\begin{equation}
    \sum_{l=0}^{\infty} d_l e^{-\epsilon l} \sim \epsilon^{-d}\,.
\end{equation}
In order to regularize this divergence one can make $d$ negative \cite{Diaz:2007an} and then continue $d$ to the positive domain. In practice, this is equivalent to saying that the total number of degrees of freedom is zero:
\begin{equation}\label{eq:trick}
    \sum_{l=0}^{\infty}d_l=0\,.
\end{equation}
Therefore, we successfully drop the first term in \eqref{eq:loggamma}. In order to pass from $\log\Gamma$ to the intermediate form one needs to apply the integral representation of $\log \Gamma(x)$:
\begin{equation}
    \log\frac{\Gamma(\mu+\nu+1)}{\Gamma(\mu+1)}=\int_0^{\infty}\frac{d\beta}{\beta}\left(\nu e^{-\beta} -\frac{e^{-\beta \mu }-e^{-\beta(\mu+\nu)}}{e^{\beta}-1}\right)\,.
\end{equation}
As a result, (\ref{eq:loggamma}) simplifies to
\begin{equation}\label{eq:intermediatestep}
    F^{\phi}_{min}=\frac{1}{2} \int_0^{\infty}\frac{d\beta}{\beta} e^{- \frac{\beta (2 + d)}{2}} \frac{ (1 + e^{\beta})^2}{(1 - e^{-\beta})^{d}}\,.
\end{equation}
By making a change of variable, $\beta \rightarrow 2\beta$, we get exactly the intermediate form  (\ref{eq:compactA}) obtained in $AdS$ up to a factor of $(-2)$. By definition, the $AdS$ one-loop free energy is related to the sphere free energy as
\begin{equation}\label{eq:FandZeta}
    F^{\phi}_{\min}=- \frac12 \tilde{\zeta}'_{\text{min.}}(0)\,,
\end{equation}
which explains the factor $(-2)$ difference. We also note that (\ref{eq:loggamma}) leads to 
\begin{equation}\label{eq:freeF}
    F^{\phi}_{\text{min.}}= \frac{1}{2}\sum_{l=0}^{\infty} d_l \log\frac{\Gamma(\frac{d}{2}+l-1)}{\Gamma(\frac{d}{2}+l+1)}= \frac{1}{\sin\left(\frac{\pi d}{2}\right)\Gamma(d+1)}\int_0^{1} du u \sin(\pi u ) \Gamma\left(\frac{d}{2}+u\right)\Gamma\left(\frac{d}{2}-u\right)\,.
\end{equation}
In Appendix \ref{app:Proof} we show that the same result can be obtained directly from the intermediate form, i.e. the $AdS$ result suffices to reproduce \eqref{eq:freeF} and there is no 'information loss' in going to the intermediate form. Then, the generalized sphere free energy  $\tilde{F}^{\phi}=-\sin(\tfrac{\pi d}{2})F_\phi$ is \cite{Giombi:2015haa,Giombi:2014xxa}:
\begin{equation}\label{eq:FA1}
    \tilde{F}^{\phi}_{\text{min.}}=\frac{1}{\Gamma(d+1)}\int_0^1 du \sin(\pi u) \Gamma\left(\frac{d}{2}-u\right)\Gamma\left(\frac{d}{2}+u\right)\,.
\end{equation}
Finally, we have shown that the (generalized) sphere free energy of the free scalar field results from the one-loop determinant in the minimal Type-A higher-spin theory:
\begin{equation}\label{genpoleF}
    -\sin\left(\frac{\pi d }{2}\right) F^{\phi}_{\text{min.}} = \tilde{F}^{\phi}_{\text{min.}} = \frac{1}{2}\sin\left(\frac{\pi d }{2}\right) \tilde{\zeta}'_{\text{min.}}(0) \,,
\end{equation}
which completes the proof. Despite the fact that our proof requires $d$ not to be an even integer, the final result smoothly extrapolates to $d=2k$, where there are poles that correspond to the $a$-anomaly. This extends the proof to even dimensions as well.

\subsection{Matching Critical Vector Model}\label{sec:matchingCritical}
Let us consider the case of the duality between the critical $O(N)$ vector model and the (non)-minimal Type-A theory where the scalar field is quantized with $\Delta=2$ ($\tilde{\nu}_{\phi}=2-\frac{d}{2}$) boundary condition. It is clear the we just need to add to $\zeta'_{\text{n.-m.}}(0)$ or $\zeta'_{\text{min.}}(0)$ the difference that is due to the change of boundary conditions for the scalar field. In this case, we see that  $\tilde{\nu}_{\phi}=-\nu_{\phi}$. As we consider the modified zeta function (\ref{eq:modzetaA}), the exponential $\exp(-\beta \nu)$ will change sign. Also, it is clear, see Appendix \ref{app:deficit}, that the deficit that can be missing from $\zeta'(0)$ due to the modified zeta-function, is absent thanks to $\tilde{\nu}_{\phi}=-\nu_{\phi}$. Repeating the procedure above, we obtain
\begin{equation}\label{eq:UV-IR}
    \delta \tilde{\zeta}'_{\phi}(0) =  \tilde{\zeta}'_{\frac{d}{2}-2,0}(0)-\tilde{\zeta}'_{2-\frac{d}{2},0}(0)  =\int_0^{\infty} \frac{ (1+e^{\beta})\left( e^{2\beta}-e^{\beta(d-2)} \right)}{\beta(e^{\beta}-1)^{d+1}}\,.
\end{equation}
This is the intermediate form that after using the same regularization as on the CFT side will give the difference between the values of the generalized sphere free energy for the free and interacting $O(N)$ vector models:
\begin{equation}
    \delta \tilde F= \tilde F_{IR}-\tilde F_{UV} = \frac{1}{\Gamma(d+1)}\int_0^{d/2-2} u \sin(\pi u)\Gamma\left(\frac{d}{2}-u\right)\Gamma\left(\frac{d}{2}+u\right)du\label{deltaF}\,.
\end{equation}
Therefore, we come to the conclusion that 
\begin{equation}\label{eq:Finteract}
    \delta F = -\frac12{\delta \tilde{\zeta}'_{\phi}}(0)
\end{equation}
Indeed, we can get (\ref{eq:Finteract}) from the CFT side through an intermediate formula which is minus one half of (\ref{eq:UV-IR}). To be more explicit,
\begin{equation}
\begin{split}
    \delta F &= \frac{1}{2}\sum_{l=0}^{\infty} d_l \log \frac{\Gamma(l+2)}{\Gamma(l+d-2)}= \frac{1}{2}\sum d_l \int_0^{\infty} \frac{d\beta}{\beta}\left((4-d)e^{-\beta}-\frac{e^{-\beta\left(l+d-3\right)}-e^{-\beta\left(l+1\right)}}{e^{\beta}-1}\right)\\
    &=-\frac{1}{2}\int_0^{\infty} \frac{ (1+e^{\beta})\left( e^{2\beta}-e^{\beta(d-2)}\right)}{\beta(e^{\beta}-1)^{d+1}}\,.
    \end{split}
\end{equation}
The same procedure as in Appendix \ref{app:Proof} allows one to relate the intermediate form to \eqref{deltaF}.

\section{Discussion and Conclusions}
\label{sec:conclusions}
Our main result is the derivation of the (generalized) sphere free energy $\tilde F$ of a free scalar field as a one-loop effect in the minimal Type-A higher-spin theory. Along the way we had to prove that the $\log$-divergence, which is given by $\zeta_{\text{HS}}(0)$, vanishes identically both in the minimal and non-minimal Type-A theories, as well as $\zeta'_{\text{HS}}(0)=0$ in the non-minimal Type-A theory. The main goal was to extend this result to all integer dimensions as well as to fractional ones. Also, we reproduced the $O(1)$ corrections to the (generalized) sphere free energy in the large-$N$ critical $O(N)$ vector model, which should be dual to the minimal Type-A theory with $\Delta=2$ boundary conditions for the scalar field. This supports the conjecture that the AdS/CFT duality may extend to fractional dimensions at least for some of the models and some of the observables that are well-defined in non-integer dimensions.

It would be interesting to extend the results of this paper to other models. For example, it should be possible to show directly in $AdS_{d+1}$ that the generalized sphere free energy of higher-spin duals of $\square^k \phi=0$ free CFT's should follow 
\begin{align}
    \tilde{F}&=\frac{1}{\Gamma (d+1)}{\int_0^{\Delta -\frac{d}{2}} u \sin (\pi  u) \Gamma \left(\frac{d}{2}-u\right) \Gamma \left(\frac{d}{2}+u\right) \, du}\,, && \Delta=\frac{1}{2} (d-2 k)\,,\quad k=1,2,...\,,
\end{align}
which is in accordance with the values for integer $d$ computed in \cite{Brust:2016gjy,Brust:2016xif}. Another motivation comes from the Type-B puzzle that has been observed in \cite{Giombi:2013fka,Pang:2016ofv,Gunaydin:2016amv,Giombi:2016pvg}. The Type-B is the higher-spin theory that is supposed to be dual to free fermion in generic dimension or to the Gross-Neveu model for $2<d<4$. The $AdS$ one-loop determinant gives the $a$-anomaly coefficient of the free fermion for $d$ even. However, the computations for $d$ odd results in a sequence of numbers that do not match the sphere energy of the free fermion, but still can be deduced from the change of the $F$-energy under a double-trace deformation \cite{Gunaydin:2016amv,Giombi:2016pvg}. It would be interesting to see what happens in fractional dimensions with the Type-B theory.

Let us note that, as was noted in \cite{Camporesi:1994ga}, there is some relation between the zeta-function of Laplace operator on the sphere $S^{d+1}$ and on the hyperbolic space $\mathbb{H}^{d+1}$. This relation is obtained by choosing certain contour for the integral over the spectral parameter $\lambda$. As a result the sum over the residues gives the zeta-function on $S^{d+1}$, but there are other contributions as well. It would be advantageous to make the relation between the computations on sphere and hyperbolic spaces more direct in the higher-spin case. It is striking that the form for the zeta-function we obtained on the AdS side is very close to the one for the Casimir energies in \cite{Giombi:2014yra}. In this regard let us mention that it is sometimes possible \cite{Giombi:2014yra} to massage the $AdS$ one-loop computations (the Casimir energy in the case of \cite{Giombi:2014yra}) in such a way that the formally divergent sum over the spins is of the same form as on the CFT side. In this case, it is clear that one needs to use coherent regularizations on the two sides of the duality. Our computation follows the same strategy and ends up on the convenient intermediate form of the one-loop result that can be directly matched with the CFT one \cite{Klebanov:2011gs,Giombi:2014xxa}. Mostly for technical reasons we used a modified regularization, see also \cite{Bae:2016rgm,Bae:2016hfy}, which leads to a certain deficit for representations whose character is not an even function of $\beta$. Fortunately, this issue does not affect the computation for the Type-A theories (after the summation over spins is performed).\footnote{The deficit at order $z$ is just the leftover of $P_{\nu,s}$ without the part including $c_n^+$ in \cite{Gunaydin:2016amv}. This allows us to conclude that the modified zeta function works well for arbitrary dimension since $\mathcal{O}(z^2)$ is irrelevant at one loop.}

Let us also briefly mention other results that support the conjecture that at least the higher-spin/vector model duality should work in a continuous range of dimensions. The scalar cubic coupling in the higher-spin theory turns out to be extremal in $AdS_4$ \cite{Sezgin:2003pt,Giombi:2009wh} and vanishing of the coupling constant near $AdS_{4}$ for the dual of the bosonic vector-model \cite{Bekaert:2014cea} and near $AdS_3$ for the dual of the Gross-Neveu model \cite{Skvortsov:2015pea} is properly compensated by the divergence of the bulk integral as a function of $d$. In principle, it should also be possible to compute the anomalous dimensions of the higher-spin currents in the critical vector and Gross-Neveu models that are well-defined for fractional dimensions. On the CFT side the anomalous dimensions of the higher-spin currents are known up to the $1/N^2$ order, \cite{Lang:1992zw,Muta:1976js,Giombi:2016zwa,Skvortsov:2015pea,Giombi:2016hkj,Giombi:2017rhm,Manashov:2016uam,Manashov:2017xtt} and up to $\epsilon^4$ for the bosonic vector model in $4-\epsilon$ expansion \cite{Wilson:1973jj,Braun:2013tva,Derkachov:1997pf,Manashov:2017xtt}.


\section*{Acknowledgments}
\label{sec:Aknowledgements}
We would like to thank Euihun Joung, Sebastian Konopka, Shailesh Lal and Tomas Prochazka for the very useful comments. We are grateful to Simone Giombi for a very useful discussion of the earlier draft. T.T. is grateful to Ivo Sachs, Igor Bertan and Katrin Hammer for useful discussions. The work of E.S. was supported in part  by the Russian Science Foundation grant 14-42-00047 
in association with Lebedev Physical Institute and by the DFG Transregional Collaborative 
Research Centre TRR 33 and the DFG cluster of excellence ``Origin and Structure of the Universe".

\begin{appendix}
\renewcommand{\thesection}{\Alph{section}}
\renewcommand{\theequation}{\Alph{section}.\arabic{equation}}
\setcounter{equation}{0}\setcounter{section}{0}

\section{From Intermediate to Final Form}
\label{app:Proof}
As a result of the $AdS$ computation we arrived at the intermediate form (\ref{eq:intermediatestep}), which can easily be seen to arise in the computation of the determinant on the CFT side. Let us now show how to reach the (generalized) sphere free energy $F_{\phi}$ in its final form. In order to compute the $\beta$-integral we use 
\begin{equation}
    \frac{1}{\beta}=\frac{1}{2}\left(\frac{1}{1-e^{-\beta}}\int_0^1 du e^{-u\beta} - \frac{1}{1-e^{\beta}}\int_0^1 du e^{u \beta} \right)\,.
\end{equation}
This allows for an analytic evaluation of the $\beta$ integral. One obtains
\begin{equation}\label{eq:doublepair}
    F_{\text{min.}}^{\phi} = \frac{\Gamma(-d)}{4} \int_0^1 du (d + 4 (-1 + u) u)\left(\frac{\Gamma\left(-1+\frac{d}{2}+u\right)}{\Gamma\left(1-\frac{d}{2}+u\right)}+\frac{\Gamma\left(\frac{d}{2}-u\right)}{\Gamma\left(2-\frac{d}{2}-u\right)} \right)
\end{equation}
After some straightforward algebra (\ref{eq:doublepair}) can be shown to split in two parts, the first one we can bring to the form of (for $\Delta=\frac{d}{2}-1$) \cite{Diaz:2007an,Giombi:2014xxa}:
\begin{equation}
\begin{split}
    F_{\Delta} &= \Gamma(-d) \int_0^{\Delta-\frac{d}{2}} du u \left[\frac{\Gamma(\frac{d}{2}-u)}{\Gamma(1-u-\frac{d}{2})}-\frac{\Gamma(\frac{d}{2}+u)}{\Gamma(1+u-\frac{d}{2})}\right] \\
    &= - \frac{1}{\sin(\frac{\pi d}{2})\Gamma(d+1)}\int_0^{\Delta-\frac{d}{2}} du u \sin(\pi u ) \Gamma\left(\frac{d}{2}+u\right)\Gamma\left(\frac{d}{2}-u\right)\,,
    \end{split}
\end{equation}
where the result for the free scalar field corresponds to $\Delta=\frac{d}{2}-1$. The second part has the form 
\begin{equation}
    \varpi = \frac{1}{4\Gamma(d)\sin\left(\frac{\pi d}{2}\right)}\int_0^1 (1-2u) \sin(\pi u ) \Gamma\left(-1+\frac{d}{2}+u\right)\Gamma\left(\frac{d}{2}-u\right) \,.
\end{equation}
However, this extra term vanishes due to the anti-symmetry of the integrand around $u=1/2$. This shows that
\begin{equation}
    F_{\text{min.}}^{\phi} = \frac{1}{2}\int_0^{\infty} \frac{d\beta}{\beta}\frac{e^{-\beta(2+d)/2} (1+e^{\beta})^2}{(1-e^{-\beta})^d} = \frac{-1}{\Gamma(d+1)\sin\left(\frac{\pi d}{2}\right)} \int_0^1 du\, u \sin(\pi u )\Gamma\left(\frac{d}{2}-u\right)\Gamma\left(\frac{d}{2}+u\right)\notag
\end{equation}
\section{Modified Zeta Function}
\label{app:modified}
In this Appendix we elaborate on the properties of the modified zeta-function we introduced in Section \ref{sec:TypeA}. It follows from the definition that the value of $\zeta_{\Delta,s}(0)$ is unaffected, which is illustrated in \ref{app:zeta}. The value of $\zeta'_{\Delta,s}(0)$ differs in general from its true value. Fortunately, $\zeta'(0)$ is still the same for  for the spectrum of (non)-minimal Type-A, which is studied in \ref{app:deficit}. It is also shown there that there is no deficit for the difference between the scalars with $\Delta=d-2$ and $\Delta=2$ boundary conditions.

\subsection{Zeta}
\label{app:zeta}
From (\ref{eq:modzetaA}), one can easily obtain the full zeta in various odd dimensions with the help of analytical continuation to the Lerch transcendent and then set $z\rightarrow0$. For example,
{\footnotesize
\begin{align*}
    d=3&: &&\tilde{\zeta}_{\nu,s}=\frac{(2s+1)(-17-40s-40s^2+240\nu^4-120(\nu+2s \nu)^2)}{5760}\\
    d=5&: &&\tilde{\zeta}_{\nu,s}=-\frac{(1+s)(2+s)(3+2s)(-1835-2142s-714s^2-1260(3+2s)^2\nu^2+5040(5+2s(s+3))\nu^4-6720\nu^6)}{29030400}
\end{align*}}
It is easy to see that these polynomials in $\nu$ and $s$ are exactly the zeta function for Type-A in \cite{Giombi:2014iua}, see also \cite{Gunaydin:2016amv}. Therefore, there is no deficit at $z^0$ order, i.e
\begin{equation}
   \tilde{\zeta}_{\nu,s}(0) -  \zeta_{\nu,s}(0)=0 + \mathcal{O}(z)\,.
\end{equation}
This explains how we can get all the correct $\tilde{\zeta}_{d,s}(0)$ for individual spins in general odd dimensions.  There are many results on zeta-function at $d=3$, see e.g. \cite{Camporesi:1993mz,Giombi:2013fka,Giombi:2014iua}. Let us illustrate that the modified zeta-function is solid enough to obtain these results. The spin factor in $d=3$ is 
\begin{equation}
    g^A_3(s)= 2s+1
\end{equation}
Together with $\nu=s-\frac{1}{2}$, \eqref{eq:modzetaA} becomes
\begin{equation*}
    \tilde{\zeta}^A_3(z) = -\frac{(2s+1)}{3!\Gamma(2z)} \int_0^{\infty} d\beta \sum_{l=0}^{\infty} e^{-\beta(s-\frac{1}{2})}\beta^{2z-1} e^{-\beta\left(\frac{1}{2}+l\right)}\left(\frac{1}{2}+l\right)
    \left(\left(\frac{1}{2}+s\right)^2-\left(\frac{1}{2}+l\right)^2\right)\,.
\end{equation*}
Now we can sum over $l$ and obtain
\begin{equation*}
    \tilde{\zeta}^A_{3,s}(z)=\frac{1}{12\Gamma(2z)}\int_0^{\infty} d\beta \frac{\beta^{2z-1} e^{-\beta(s-1)} (1 + e^{\beta}) (1 + 2 s) (s (1 + s) + e^{2 \beta} s (1 + s) - 
    2 e^{\beta} (3 + s + s^2))}{(-1 + e^{\beta})^4}
\end{equation*}
In order to get to the actual numbers one needs to plug $s=0,1,2,3,...$ then use the trick of analytical continuation via the Hurwitz-Lerch zeta function \cite{Giombi:2013fka,Giombi:2014iua}. For example,
\begin{equation}
    \tilde{\zeta}^A_{3,s}(0)=\left\{-\frac{1}{180},-\frac{11}{60},-\frac{181}{36},-\frac{6097}{180},...\right\}
\end{equation}
Note that, after the continuation to the Hurwitz-Lerch transcendent, there will be another $\Gamma(2z)$ function in the nominator. This will cancel $1/\Gamma(2z)$ factor in the modified zeta function. Therefore, the modified zeta-function reproduces the correct result, which is expected.

\subsection{Deficit}
\label{app:deficit}
As we already explained in Section \ref{sec:TypeA}, we changed the regularization prescription. As a result the values of $\zeta'_{\Delta,s}(0)$ might be different from the correct ones for individual fields. It was noted in \cite{Bae:2016rgm} that the deficit vanishes for certain representations (with even character). In particular, the deficit is absent for (non)-minimal Type-A theory. The purpose of this Section is to quantify the deficit for a number of cases.

For example, let us take the scalar field in $d=3$. The zeta-prime can be derived by calculating $\zeta(z)$ at $z$ order:
\begin{equation}\label{eq:deficit1}
    \small \zeta_{3,0}(z)=\frac{\zeta(-3+2z)}{6}+\frac{\zeta(-2+2z)}{4}+\frac{\zeta(-1+2z)}{12} = -\frac{1}{180}+\left(\frac{1}{72}-\frac{\log A}{6}+\frac{\zeta'(-3)}{3}+\frac{\zeta'(-2)}{2}\right)z + \mathcal{O}(z^2) 
\end{equation}
One can already notice that there is a deficit between the value of $\tilde{\zeta}'^A_{3,0}(0)$ that is evaluated by the standard zeta function and (\ref{eq:deficit1}). This was also discussed in Appendix (B.1) of \cite{Bae:2016rgm}, when the authors use characters to evaluate $\tilde{\zeta}'(0)$ for different fields. Let us have a look at the deficit in $d=3$ and $d=5$ as to observe the general pattern.
\subsubsection{d=3}
The result before sending $z$ to $0$ for the modified zeta function is
\begin{equation}
\begin{split}
   \tilde{\zeta}^3_{\nu,s}(z)= &\frac{(2s+1)}{24}\Big[\nu\left((1+2s)^2-4\nu^2\right)\zeta(2z,\nu+\frac{1}{2})+4\zeta(-3+2z,\nu+\frac{1}{2})\\
    &-12\nu \zeta(-2+2z,\nu+\frac{1}{2})+(-1-4s(1+s)+12\nu^2)\zeta(-1+2z,\nu+\frac{1}{2})\Big]\,.
    \end{split}
\end{equation}
In order to compute the zeta-prime, one just needs to take the $z$ derivative and set $z=0$:
\begin{equation}
    \begin{split}
   \tilde{\zeta}'^3_{\nu,s}(0)= &\frac{(2s+1)}{12}\Big[\nu\left((1+2s)^2-4\nu^2\right)\zeta'(0,\nu+\frac{1}{2})+4\zeta'(-3,\nu+\frac{1}{2})\\
    &-12\nu \zeta'(-2,\nu+\frac{1}{2})+(-1-4s(1+s)+12\nu^2)\zeta'(-1,\nu+\frac{1}{2})\Big]\,.
    \end{split}
\end{equation}
We then follow the procedure in \cite{Bae:2016rgm} to find the deficit. First, we set $\nu=0$ and obtain
\begin{equation}
    \tilde{\zeta}'^3_{0,s}(0)=\frac{(2s+1)}{12}\left(4\zeta'(-3,\frac{1}{2})-(2s+1)^2\zeta'(-1,\frac{1}{2})\right)\,.
\end{equation}
Recall that for the standard zeta-prime in $d=3$, see \cite{Giombi:2013fka,Giombi:2014iua}, we have
\begin{equation}
    \zeta'^3_{0,s}(0)=\frac{2s+1}{3}\left(c_3 + \left(s+\frac{1}{2}\right)^2c_1\right)\,.
\end{equation}
We note that 
\begin{equation}\label{eq:identity1}
    \zeta'(-n,\frac{1}{2})=(-)^{\frac{n+1}{2}}c_n, \qquad \text{where} \quad c_n=\int_0^{\infty}du \frac{2u^n \log u}{e^{2\pi u}+1}\,.
\end{equation}
Therefore, $\zeta'^3_{0,s}(0)$ and $\widetilde{\zeta}'^3_{0,s}(0)$ do match. Then, we consider the $\nu$ derivatives for each of the zetas:
\begin{equation}
\begin{split}
    \partial_{\nu}\tilde{\zeta}'^3_{\nu,s}(0)=&\frac{(2s+1)}{12}\Bigg(\left((2s+1)^2-12\nu^2\right)\zeta'(0,\nu+\frac{1}{2})-12\zeta'(-2,\nu+\frac{1}{2})+24\nu \zeta'(-1,\nu+\frac{1}{2})\\
    &+\nu((2s+1)^2-4\nu^2)\partial_{\nu}\zeta'(0,\nu+\frac{1}{2})+4\partial_{\nu}\zeta'(-3,\nu+\frac{1}{2})-12\nu\partial_{\nu}\zeta'(-2,\nu+\frac{1}{2})\\
    &+(-1-4s(s+1)+12\nu^2)\partial_{\nu}\zeta'(-1,\nu+\frac{1}{2})+\nu((2s+1)^2-4\nu^2)\partial_{\nu}\zeta'(0,\nu+\frac{1}{2})\Bigg)
    \end{split}
\end{equation}
\begin{equation}\label{zetadev3}
    \partial_{\nu}\zeta'^3_{\nu,s}(0)=\frac{(2s+1)}{3}\left(\frac{\nu^3}{2}+\frac{\nu}{24}+\nu\left(\left(s+\frac{1}{2}\right)^2-\nu^2\right) \psi(\nu+\frac{1}{2})\right)
\end{equation}
Next, using the identities for Hurwitz zeta function
\begin{equation}\label{eq:identity2}
    \partial_{\nu}\zeta(s,\nu)=-s\zeta(s+1,\nu), \qquad \partial_{\nu}\zeta'(0,\nu)=\psi(\nu)
\end{equation}
we can reduce the $\nu$ derivative of the modified zeta-prime to
\begin{equation}\label{modzetadev3}
     \partial_{\nu}\tilde{\zeta}'^3_{\nu,s}(0)=\frac{(2s+1)\nu((2s+1)^2-4\nu^2)\psi(\nu+\frac{1}{2})}{12}
\end{equation}
Subtracting (\ref{modzetadev3}) and (\ref{zetadev3}) together, then integrating over $\nu$, one obtains the deficit for individual fields at order $z$:
\begin{equation}
    \delta \zeta'_{\nu,s}(0)=\tilde{\zeta}'_{\nu,s}-\zeta'_{\nu,s}=-\frac{(2s+1)(\nu^2+6\nu^4)}{144}
\end{equation}
Since the deficit is an even function of $\nu$, we can compute the difference between the scalars with $\Delta=d-2,2$ boundary conditions using the modified zeta function thanks to  $\delta\zeta'_{d-2,0}-\delta\zeta'_{2,0}=0$.
Using the cut-off $e^{-\epsilon(s+\frac{d-3}{2})}$, one can sum over either all spins or even spins and observe that the deficit does vanish:
\begin{equation}
    \sum_{s} \delta \zeta'_{\nu,s}(0)=0\,.
\end{equation}
Therefore, the deficit is absent both for the non-minimal and minimal Type-A theories at order $z$, which is what we need for $\zeta'_{\text{HS}}(0)$. 
\subsubsection{d=5}
In higher dimensions, there is another useful identity that we illustrate on the example of $d=5$. Following the procedure outlined above, we obtain
{\allowdisplaybreaks{\begin{equation*}
\begin{split}
    \small\tilde{\zeta}'^5_{\nu,s}(0)=&\frac{(1 + s) (2 + s) (3 + 2 s)}{5760}\Bigg[-16\zeta'(-5,\nu+\frac{3}{2})+8 \nu (-3 (5 + 2 s (3 + s)) + 20 \nu^2)\zeta'(-2,\nu+\frac{3}{2})\\
    &+80\nu \zeta'(-4,\nu+\frac{3}{2})+(-(3 + 2 s)^2 + 24 (5 + 2 s (3 + s)) \nu^2 - 80 \nu^4)\zeta'(-1,\nu+\frac{3}{2})\\
    &+\nu (-1 + 4 \nu^2) (-9 - 4 s (3 + s) + 4 \nu^2)\zeta'(0,\nu+\frac{3}{2})+8 (5 + 2 s (3 + s) - 20 \nu^2)\zeta'(-3,\nu+\frac{3}{2})\Bigg]\,.
    \end{split}
\end{equation*}}}%
Setting $\nu=0$ we arrive at 
\begin{equation*}
\begin{split}
    \tilde{\zeta}'^5_{0,s}(0)=&\frac{(1 + s) (2 + s) (3 + 2 s)}{5760}\Bigg[-16\zeta'(-5,\frac{3}{2})+8 (5 + 2 s (3 + s))\zeta'(-3,\frac{3}{2})-(3 + 2 s)^2\zeta'(-1,\frac{3}{2})\Bigg]\,.
    \end{split}
\end{equation*}
We massage the formula above as to be able to compare $\zeta'(-k,\tfrac12)$ with $c_n$, which can be done with the help of
\begin{equation}\label{eq:identity3}
    \zeta(s,\nu)=\zeta(s,\nu+m)+\sum_{n=0}^{m-1}\frac{1}{(n+\nu)^s}
\end{equation}
We arrive at
\begin{equation}\label{modzetaprime5}
\begin{split}
    \tilde{\zeta}'^5_{0,s}(0)=&-\frac{(1 + s) (2 + s) (3 + 2 s)}{5760}\Bigg[-16\zeta'(-5,\frac{1}{2})+8 (5 + 2 s (3 + s))\zeta'(-3,\frac{1}{2})-(3 + 2 s)^2\zeta'(-1,\frac{1}{2})\Bigg]\,,
    \end{split}
\end{equation}
which can be compared with the standard zeta-prime: \begin{equation}\label{zetaprime5}
    \zeta'^5_{0,s}(0)=-\frac{(1 + s) (2 + s) (3 + 2 s)}{360}\left(c_5+c_3\left(\frac{1}{4}+\left(s+\frac{3}{2}\right)^2\right)+\frac{c_1}{4}\left(s+\frac{3}{2}\right)^2\right)\,.
\end{equation}
Using the identity (\ref{eq:identity1}), it is easy to realize that  (\ref{modzetaprime5}) and (\ref{zetaprime5}) are the same. Next, one can proceed as in the previous Section and get
\begin{equation}
    \delta \zeta'^5_{\nu,s}=-\frac{(s+1)(s+2)(2s+3)\nu^2(107+580\nu^2-240\nu^4+120s(1+6\nu^2)+40s^2(1+6\nu^2))}{691200}\,.
\end{equation}
The sum over all (even) spins can be found to vanish, which guarantees that the deficit does not contribute to the zeta-prime of the (non)-minimal Type-A. Also the deficit is an even function of $\nu$ and therefore the difference due to $\Delta=d-2,2$ boundary conditions for the scalar field is also free of any deficit.

Let us note that the deficit has already appeared in implicit form in the literature. It is the leftover of $P_{\nu,s}$ in \cite{Gunaydin:2016amv} without the part including $c_n^+$, see also \cite{Giombi:2014iua} where the same structures are present but in different notation.

\end{appendix}

\setstretch{0.93}

\providecommand{\href}[2]{#2}\begingroup\raggedright\endgroup

\end{document}